\documentclass[%
 reprint,
 superscriptaddress,
 amsmath,amssymb,
 aps,
 prx,
 nofootinbib,
]{revtex4-2}

\usepackage{graphicx}    
\usepackage{dcolumn}     
\usepackage{bm}          
\usepackage{booktabs}    
\usepackage{xcolor}      
\usepackage{microtype}   

\usepackage{hyperref}

\usepackage[caption=false]{subfig}

\begin{document}

\preprint{APS/123-QED}

\title{Real-Time Detection of Charge Jumps in Superconducting Qubits\\
       with a Convolutional Neural Network}

\author{Daniel Gaytan-Villarreal}
\thanks{Corresponding author: jgaytanv@andrew.cmu.edu}
 \affiliation{Department of Physics, Carnegie Mellon University, Pittsburgh, PA 15213, USA}

\author{Peter Meiring}
 \affiliation{Department of Physics, Carnegie Mellon University, Pittsburgh, PA 15213, USA}

\author{Daniel Baxter}
 \affiliation{Quantum Division, Fermi National Accelerator Laboratory, Batavia, IL 60510, USA}
 \affiliation{Department of Physics \& Astronomy, Northwestern University, Evanston, IL 60208, USA}

\author{Daniel Bowring}
 \affiliation{Quantum Division, Fermi National Accelerator Laboratory, Batavia, IL 60510, USA}

\author{Grace Bratrud}
 \affiliation{Department of Physics \& Astronomy, Northwestern University, Evanston, IL 60208, USA}
 \affiliation{Quantum Division, Fermi National Accelerator Laboratory, Batavia, IL 60510, USA}

\author{Matteo Cremonesi}
 \affiliation{Department of Physics, Carnegie Mellon University, Pittsburgh, PA 15213, USA}

\author{Giuseppe Di Guglielmo}
 \affiliation{Fermi National Accelerator Laboratory, Batavia, IL 60510, USA}

\author{Grace Wagner}
 \altaffiliation{now at Colorado School of Mines, Golden, CO 80401, USA}
 \affiliation{Quantum Division, Fermi National Accelerator Laboratory, Batavia, IL 60510, USA}

\author{Bowen Xiao}
 \affiliation{Department of Physics, Carnegie Mellon University, Pittsburgh, PA 15213, USA}

\date{\today}

\begin{abstract}
Ionizing radiation from cosmic rays and gammas can induce discontinuous jumps in the
environmental charge of superconducting qubits (charge jumps), causing correlated errors
that challenge fault-tolerant quantum computing while simultaneously providing a detection
signature for quantum sensing applications. Current detection methods operate offline,
introducing latency incompatible with in-the-loop qubit control. In this paper, an online
detector of charge jumps for superconducting qubits, based on a dilated causal
convolutional neural network (DCCNN) designed for in-the-loop deployment on the Quantum
Instrumentation Control Kit (QICK) platform, is presented. The network is trained on
synthetic Ramsey tomography scans generated from qubit templates measured at the
Northwestern Experimental Underground Site (NEXUS) at Fermilab, and translated to FPGA
firmware via \textsc{hls4ml} with $\texttt{ap\_fixed}\langle 16,6 \rangle$ quantization,
reaching a per-inference latency of $32.0\,\mu$s on the Zynq UltraScale+ RFSoC ZCU216,
with the compiled firmware reproducing the software model's detection efficiency to
within $0.002$. At this operating point the DCCNN matches the detection efficiency of the
established offline $\chi^2$ algorithm ($0.843 \pm 0.022$ vs.\ $0.868 \pm 0.007$ on
$|\Delta q| \in [0.1, 0.5]\,e$ at matched false-positive rate), while requiring no
per-qubit hyperparameter tuning. This shifts charge-jump detection from a post-hoc
diagnostic to a control-loop primitive, enabling adaptive protocols that respond to
radiation-induced events in situ, with applications to quantum-computing error mitigation
and to the use of superconducting qubits as particle detectors.
\end{abstract}

\maketitle

\section{Introduction}

\subsection{Qubits as new physics detectors}

Superconducting qubits are anharmonic oscillators formed by a Josephson junction shunted
by a large capacitor, operating as artificial atoms at millikelvin temperatures and with
transition frequencies in the 4--8~GHz range. Modern qubit design achieves exponential
suppression of sensitivity to low-frequency charge noise by operating in a regime where
the Josephson energy greatly exceeds the charging energy ($E_J/E_C\gg1$), enabling
coherence times over hundreds of microseconds and forming the basis for practical quantum
technologies. Qubit sensitivity to single-photon and charge perturbations, while a
challenge for quantum computing, simultaneously enables their use as quantum sensors for
fundamental physics searches \cite{danilin2024quantumsensingtunablesuperconducting,Bonomo:2026num,Ramanathan_2026,Dixit_2021,Fink_2024,Linehan_2025,linehan2024listeningnewphysicsquantum,celi2026measuringquasiparticledynamicsparticle}.

The interaction of ionizing radiation with the substrate of a superconducting qubit device
produces liberated charge-hole pairs and phonons that can travel macroscopic
distances~\cite{Cardani:2022blq}. Some of these free charges become trapped on impurities
and at interfaces, where they alter the local electric field environment~\cite{Wilen:2020lgg}.
Because qubits and other quantum systems are sensitive to local electric fields, this
discrete ``charge jump'' can functionally detune a qubit, resulting in a shift in qubit
frequency and a corresponding suppression of qubit coherence time. Superconducting qubits
used in QPU architectures are optimized for charge insensitivity and are therefore not able
to directly measure this effect. However, two-level systems (TLS) near even an optimized
qubit are highly charge-sensitive and can move in frequency following charge jumps, in what
is referred to as TLS scrambling~\cite{Thorbeck_2023}. This can further frustrate qubit
control, as the TLS coupling to the qubit state changes in strength depending on the
frequency shift. \\
Charge jumps are spatially correlated across length scales exceeding
600~$\mu$m and are often accompanied by transient suppression of qubit relaxation times
across centimeter-scale chips~\cite{Wilen:2020lgg}---posing a fundamental challenge for
fault-tolerant quantum computing, where error correction schemes assume uncorrelated errors.
The dynamics of charge jumps can be studied in charge-sensitive qubits via Ramsey
tomography as shifts in the qubit transition frequency. These same dynamics enable the use
of qubits as detectors of phenomena~\cite{Dixit_2021,Fink_2024,Linehan_2025,linehan2024listeningnewphysicsquantum,celi2026measuringquasiparticledynamicsparticle} beyond those described by the Standard Model of
particle physics~\cite{GLASHOW1961579, Weinberg:1967tq, Glashow:1970gm, GELLMANN1964214, Zweig:1964jf},
with charge jumps providing access to energy thresholds as low as a few eV. This dual
nature of superconducting qubits---as computational elements susceptible to environmental
disturbances and as sensitive probes of those same disturbances---motivates the development
of real-time charge jump detection algorithms that can support both quantum error
mitigation and fundamental physics searches.

\subsection{Ramsey measurements at NEXUS}

The Northwestern Experimental Underground Site (NEXUS) at Fermi National Accelerator
Laboratory provides a unique environment for studying the interaction of ionizing radiation
with superconducting qubits. Located 107 meters below the Earth's
surface in the former MINOS tunnel, NEXUS reduces the cosmic ray muon flux by over 99\%
compared to surface laboratories. A movable lead shield provides additional $4\pi$
coverage, enabling quantifiable control over the flux of gamma radiation incident on qubit
devices.

Charge tomography at NEXUS employs a Ramsey pulse sequence (X/2--Idle--X/2) applied to
weakly charge-sensitive qubits ($E_J/E_C=24$). During the idle period, the qubit state
vector phase evolves as a function of the offset charge:
\begin{equation*}
\phi(n_g) = 2\pi \epsilon(n_g)\, t_{\text{idle}} / \hbar
\end{equation*}
where $\epsilon(n_g)$ is the charge dispersion of the qubit and $n_g$ is the dimensionless
offset charge. This pulse sequence maps gate charge (modulo $1e$) onto the excited state
probability $P_1$ of the qubit, with the mapping designed to be insensitive to
quasiparticle parity. Scanning the charge bias across a range of voltages enables
calibration of offset charge values and extraction of $P_1$ with reduced sensitivity to
charge noise.

Long-time-series measurements at NEXUS have captured discontinuous jumps in offset charge
corresponding to interactions of ionizing radiation with the qubit
substrate~\cite{Bratrud:2024qnk}. The rate and spatial correlation of these charge jumps are monitored by energy-resolving detectors operating simultaneously in the same cryostat. Under minimal background
conditions with lead shielding closed, charge jump rates as low as 0.19~mHz were achieved,
with zero correlated events observed across qubits separated by more than 3~mm during 22
hours of continuous operation~\cite{Bratrud:2024qnk}.

\subsection{Charge jump detection with the \texorpdfstring{$\chi^2$}{chi-squared} algorithm}

Previously, the approach groups have employed for identifying charge jumps in Ramsey
tomography data employs a point-by-point analysis that fits each charge sweep to a
template function and monitors for discontinuous changes in the fitted offset charge using
a running $\chi^2$ window, hereafter referred to as the $\chi^2$ algorithm. As described in
Ref.~\cite{Bratrud:2024qnk}, for each prefix of length $n$ within a scan (with
$n \in [1, N]$ and $N$ the total scan length), the best-fit template phase $\theta_n^{\min}$
is determined by minimizing the reduced $\chi^2$ between the data and the qubit-specific
template,
\begin{equation*}
\chi^2_n(\theta) = \frac{1}{n} \sum_{i=1}^{n}
  \frac{(x_i - \hat{x}(\theta))^2}{\hat{\sigma}^2(\theta)} \,.
\end{equation*}
A charge jump within the scan introduces a discontinuity in $\theta_n^{\min}$ that causes
$\chi^2_n(\theta_n^{\min})$ to rise rapidly with each subsequent sample. A jump is
identified when this rolling value crosses a per-qubit threshold; the procedure then
resets, with the first sample above threshold treated as $i=1$ in the next iteration, and
continues until the end of the scan.

While effective for offline analysis, the cumulative nature of the $\chi^2$ algorithm
introduces a known limitation: reduced sensitivity to jumps occurring within the first
$\sim$20 samples of each scan, where insufficient pre-jump data has accumulated for the
rolling statistic to build up. Combined with the per-qubit threshold tuning required by
the algorithm, this motivates a search for methods better suited to real-time operation.
This limitation is in addition to imposing a significant time delay on the measurement, as
described in the previous paragraph.
Convolutional neural networks offer a promising approach: once trained, inference can be
performed with fixed, predictable latency, and the network can learn to identify subtle
features in the data that distinguish genuine charge jumps from noise fluctuations.
Furthermore, modern tools for translating neural networks to FPGA firmware enable
deployment of trained models directly on qubit control hardware with nanosecond-scale
inference times~\cite{NIPS2012_c399862d,QICK:2023ujm,DiGuglielmo:2025zod}, making them
well suited to applications requiring real-time feedback, such as adaptive quantum control
protocols that condition subsequent operations on the detection of a charge event, or
triggering systems for particle physics experiments.

\section{Algorithm Design and Training}

\subsection{Proposed architecture}

CNNs have been previously proposed for direct qubit state classification because of their
spatial invariance and ability to mitigate cross-talk effects in multi-qubit
configurations \cite{lienhard2021deepneuralnetworkdiscrimination,Navarathna_2021}. This
same simplicity allows for light-weight models with low inference time, which makes them
ideal candidates for real-time signal processing \cite{jwa2022realtimeinference2dconvolutional}.
Recent implementations have extended this utility to the autonomous calibration of
spin-based semiconductor qubits, where CNNs act as online charge-state tuners by
identifying electron regimes within 2D charge stability
diagrams \cite{Yon_2025,samaha2026automaticchargestatetuning}.

Because of the sequential nature of the time-series measurements at NEXUS, two models
based on the Temporal Convolutional Network (TCN) architecture were proposed. TCNs are a
sub-type of CNNs designed for time-series and sequential data. The distinguishing
characteristics of this architecture are the addition of residual connections between
convolution blocks and two changes to the convolution operator: (1) causality, which
guarantees no information leakage from future to past; and (2) dilation, which ``skips''
a fixed number of neighboring time-steps to have a wider view of the signal without
additional parameters \cite{DBLP:journals/corr/abs-1803-01271}.

\begin{figure}[ht]
    \centering
    \includegraphics[width=\columnwidth,height=0.35\textheight,keepaspectratio]{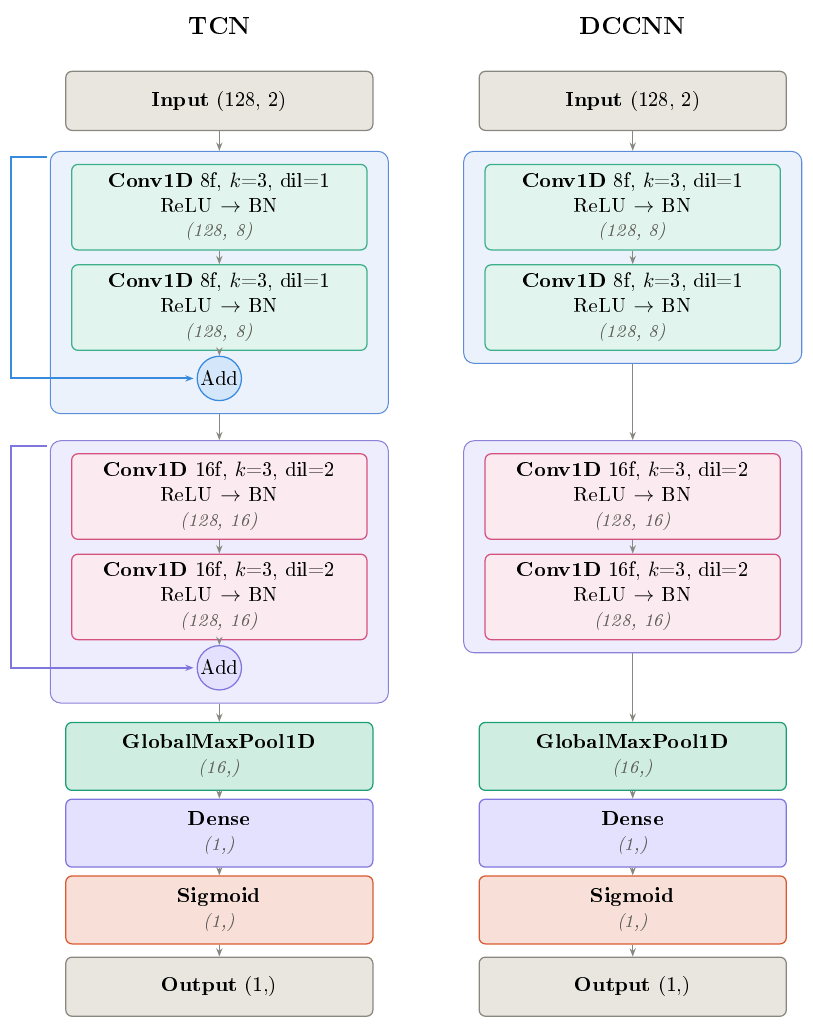}
    \caption{Proposed architectures: TCN (left) and DCCNN (right). Note that the DCCNN
        has the same architecture as the TCN minus the residual connections. Tensor shapes
        are annotated inside each block.}
    \label{fig:arch-comparison}
\end{figure}

The full proposed architectures are shown in Fig.~\ref{fig:arch-comparison}.
The left architecture is a regular TCN with two convolution blocks of two
one-dimensional convolution layers each. The first convolution block has layers with 8
filters, a kernel size of 3, and a dilation rate of 1. The second convolution block
increases the number of filters to 16 and the dilation rate to 2.

TCNs inherently utilize residual skip-connections to facilitate gradient flow. To isolate
the specific contribution of these connections to the model's detection accuracy, we also
evaluate a TCN in which the residual paths are removed, which is the right architecture
shown in Fig.~\ref{fig:arch-comparison}, while maintaining identical depth, filter counts,
and dilation rates. This model is more formally referred to as a Dilated Causal CNN (DCCNN).

Table~\ref{tab:model-size} compares the two models in terms of trainable and
non-trainable parameters. Trainable parameters comprise convolutional kernels and biases;
non-trainable parameters comprise BatchNormalization moving statistics. The TCN has 168
more trainable parameters than the DCCNN, owing to the $1 \times 1$ convolutions in its
residual skip connections, which project feature maps to match channel counts between
blocks.

\begin{table}[ht]
    \caption{Model size comparison for the TCN and DCCNN architectures.}
    \label{tab:model-size}
    \begin{ruledtabular}
    \begin{tabular}{lcc}
    \textrm{Quantity} & \textrm{TCN} & \textrm{DCCNN} \\
    \colrule
    Trainable parameters     & $1{,}721$ & $1{,}553$ \\
    Non-trainable parameters & $96$      & $96$ \\
    Total parameters         & $1{,}817$ & $1{,}649$ \\
    \end{tabular}
    \end{ruledtabular}
\end{table}

Both architectures were first implemented in Python through the Keras
library \cite{chollet2015keras}. The models were later quantized using
QKeras \cite{coelho2021qkeras,qkeras_software}.
QKeras is a quantization extension that provides drop-in functional equivalents for
standard Keras layers, enabling Quantization-Aware Training
(QAT) \cite{jacob2017quantization}, which helps maintain model performance on FPGA by
simulating low-precision arithmetic during the training
phase \cite{Chen2020Statistical,Chen2021ActNN}. QAT allows the model to learn how to
operate under precision constraints while at the same time minimizing accuracy loss due to
quantization and significantly reducing model size, making it deployable on devices with
limited memory and hardware resources.

The quantized architectures are the same as those shown in Fig.~\ref{fig:arch-comparison},
with two key changes: (1) the \texttt{Conv1D} layers are replaced by the QKeras
quantized-equivalent \texttt{QConv1D}, and (2) \texttt{GlobalMaxPool} is replaced by an
\texttt{AvgPool} block to ensure compatibility with \textsc{hls4ml}. To find the optimal
fixed-point precision for the quantized architectures, a bit-width sweep was performed
across three configurations: \texttt{ap\_fixed<6,1>}, \texttt{ap\_fixed<8,3>}, and
\texttt{ap\_fixed<16,6>}. As the architectures met the hardware and latency requirements
at all three precisions, \texttt{ap\_fixed<16,6>} was chosen, as it achieved the highest
accuracy without compromising any of the constraints. This ensured that the quantization
minimized the accuracy loss while maintaining a significantly reduced resource footprint on
the FPGA.

\subsection{Synthetic data generation}
\label{sec:Data-preproc}

To train and evaluate the proposed CNN architectures, synthetic Ramsey-style charge-sweep
scans were generated using a procedure that reproduces the simulation methodology of
Ref.~\cite{Bratrud:2024qnk}. Each scan consists of $N = 80$ samples of the qubit's transmission amplitude as a function of applied gate voltage swept linearly across
$[-0.5, +0.5]\,$V.

For each of the five qubit configurations considered in Ref.~\cite{Bratrud:2024qnk}
(Q1--Q3 in superconducting (SC) mode, Q4 in both SC and shield-open (SO) modes), a
per-qubit template $\hat{x}(\theta)$ and noise envelope $\hat{\sigma}(\theta)$ are
obtained from qubit calibration data and stored as $16\times$ linearly upsampled arrays
spanning one full period of the qubit's charge-modulation response. The upsampled
representation is used for jump injection and noise drawing, then deflated back to the
native 80-point grid.

\begin{figure}[ht]
    \centering
    \subfloat[Scan without a charge jump.\label{fig:scan-nojump}]{%
        \includegraphics[width=0.85\columnwidth]{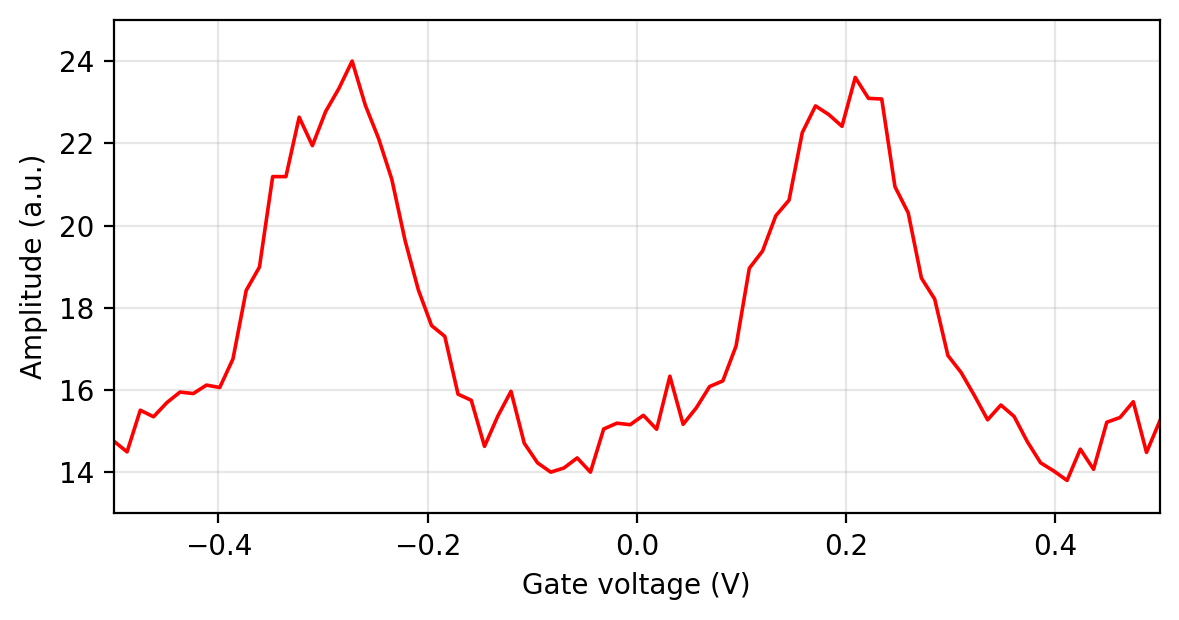}%
    }

    \vspace{0.5em}

    \subfloat[Scan containing a charge jump (dashed line) of $\Delta V = +0.200\,$V.\label{fig:scan-jump}]{%
        \includegraphics[width=0.85\columnwidth]{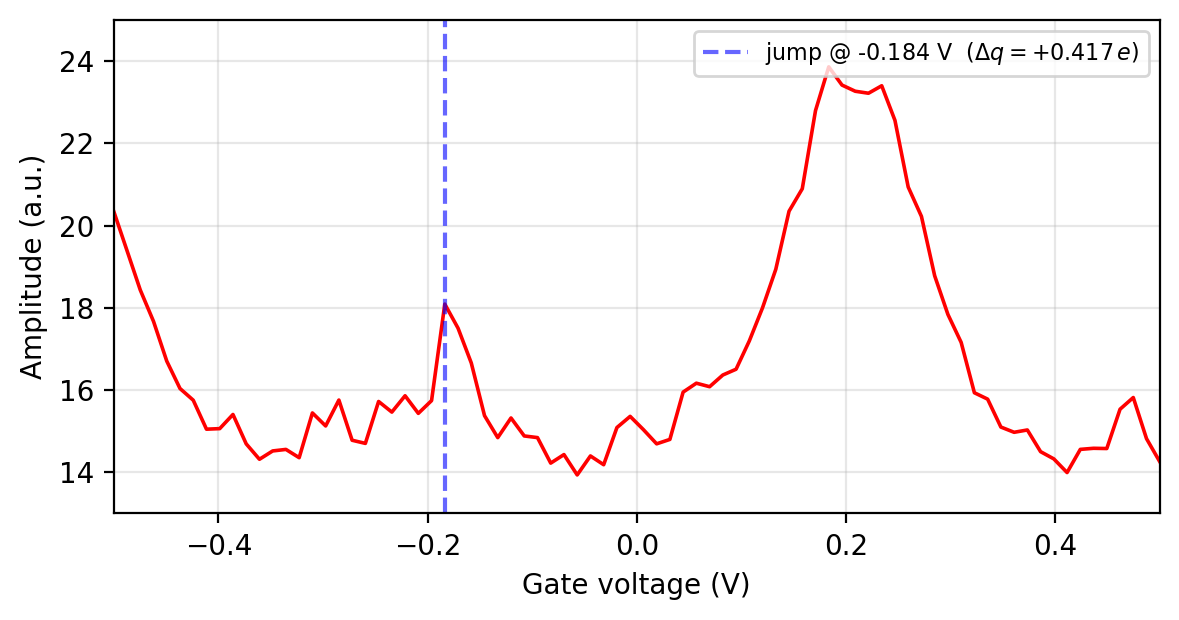}%
    }

    \caption{Example synthetic Ramsey scans generated for Qubit~1 (SC).
        Each scan covers 80 samples spanning gate voltage $[-0.5, +0.5]\,$V.
        Top: a typical no-jump scan, showing the periodic charge-modulation
        response. Bottom: a scan containing an injected charge jump
        (dashed line) of $\Delta q = +0.417\,e$ ($\Delta V = +0.200$~V) at
        gate voltage $-0.184$~V.}
    \label{fig:scan-examples}
\end{figure}

For each generated scan, the number of jumps is drawn from a Poisson distribution with
rate $\lambda = 1.1\,$mHz, matching the rate simulated in Ref.~\cite{Bratrud:2024qnk};
jump magnitudes are drawn uniformly from $[0.01, 0.5]\,e$ with random sign, and jump
positions within the scan are sampled uniformly. Scans for each qubit are generated as a
phase-continuous time-ordered sequence so that the post-event phase of scan $k$ becomes
the pre-event phase of scan $k+1$, reproducing the experimental sequence of consecutive
Ramsey acquisitions. After jump injection, per-point Gaussian noise is added using the
qubit-specific $\hat{\sigma}$ envelope. Figure~\ref{fig:scan-examples} shows representative scans
from this pipeline, comparing a jump-free signal to one with an injected discontinuity.




Each scan is converted to a two-channel input by concatenating the raw amplitude trace with its first-difference channel $\Delta x_i = x_i - x_{i-1}$ (with $\Delta x_0 = 0$). The first-difference channel exposes local sample-to-sample variation, which is the primary signature of a discontinuous phase shift.

\subsection{Training procedure}
\label{sec:training}

Both models were trained on the same fully simulated dataset of 100,000 scans generated
following the procedure of Sec.~\ref{sec:Data-preproc}, of which $\sim$33,000 contain a
charge jump. The data was split into training, validation, and test sets containing
70,000, 15,000, and 15,000 scans respectively, with the jump/no-jump proportion preserved
across splits via stratified sampling. Training used the binary cross-entropy loss (BCE)
with the Adam optimizer on a single NVIDIA A100 GPU. Keras's deterministic mode was
enabled to ensure reproducibility.

\begin{table}[ht]
    \caption{Training configuration.}
    \label{tab:training-config}
    \begin{ruledtabular}
    \begin{tabular}{ll}
    \textrm{Parameter} & \textrm{Value} \\
    \colrule
    Optimizer                              & Adam \\
    Loss                                   & BCE \\
    Maximum epochs                         & 500 \\
    Batch size                             & 256 \\
    Initial learning rate (float / QAT)    & $10^{-3}$ / $10^{-4}$ \\
    Random seed                            & 42 \\
    Early stopping patience                & 40 epochs \\
    LR reduction factor                    & 0.5 \\
    LR reduction patience                  & 7 epochs \\
    Minimum learning rate                  & $10^{-6}$ \\
    \end{tabular}
    \end{ruledtabular}
\end{table}

Training configuration is summarized in Table~\ref{tab:training-config}; both float
training and the quantization-aware retraining stage shared this configuration except for
the initial learning rate ($10^{-3}$ for float training, $10^{-4}$ for QAT retraining on
the pretrained float weights). In all cases, training was terminated by the early-stopping
criterion well before the maximum epoch count.

\section{Performance Measurements}

\subsection{Binary classification performance metrics}
\label{sec:class-perf}

Classifying whether a scan contains a charge jump is inherently a binary classification
problem, as events are split into the ``Jump'' and ``No-Jump'' classes. To evaluate both
models in a general way, standard binary classification metrics
on the held-out test set are reported: per-class precision $P$ (the fraction of predicted
jumps that are correctly identified), recall $R$ (the fraction of true jumps that are
detected), and the F1 score (the harmonic mean of precision and recall). The Jump-class
recall is equivalent to the magnitude-averaged detection efficiency.

Tables~\ref{tab:classification-float} and~\ref{tab:classification-qat} report the
per-class metrics before and after quantization, respectively. The DCCNN outperforms the
TCN on every metric in both regimes, with the clearest gap in the Jump class: in the float
model the DCCNN improves Jump precision by 4\% and Jump recall by 5\% over the TCN,
yielding a 5\% higher Jump F1 score. After quantization, the gap is preserved (5\% in
recall, 3\% in F1), with both models retaining their relative ordering. The No-Jump class
is comparatively easy for both architectures (recall $\geq 0.92$ in all cases), reflecting
the class imbalance in the training data, while Jump recall remains the limiting factor at
$\sim$0.70--0.75.

\begin{table}[ht]
    \caption{Per-class precision, recall, and F1 score for the TCN and DCCNN
        models on the held-out test set before quantization.}
    \label{tab:classification-float}
    \begin{ruledtabular}
    \begin{tabular}{lcccccc}
     & \multicolumn{3}{c}{TCN} & \multicolumn{3}{c}{DCCNN} \\
    \textrm{Class} & $P$ & $R$ & F1 & $P$ & $R$ & F1 \\
    \colrule
    No Jump & $0.86$ & $0.92$ & $0.89$ & $0.89$ & $0.94$ & $0.91$ \\
    Jump    & $0.82$ & $0.70$ & $0.75$ & $0.86$ & $0.75$ & $0.80$ \\
    \end{tabular}
    \end{ruledtabular}
\end{table}

\begin{table}[ht]
    \caption{Per-class precision, recall, and F1 score for the TCN and DCCNN
        models after quantization-aware training (\texttt{ap\_fixed<16,6>}).}
    \label{tab:classification-qat}
    \begin{ruledtabular}
    \begin{tabular}{lcccccc}
     & \multicolumn{3}{c}{TCN} & \multicolumn{3}{c}{DCCNN} \\
    \textrm{Class} & $P$ & $R$ & F1 & $P$ & $R$ & F1 \\
    \colrule
    No Jump & $0.87$ & $0.96$ & $0.91$ & $0.89$ & $0.96$ & $0.92$ \\
    Jump    & $0.90$ & $0.69$ & $0.78$ & $0.90$ & $0.74$ & $0.81$ \\
    \end{tabular}
    \end{ruledtabular}
\end{table}

\subsection{Efficiency per qubit}
\label{sec:eff-per-qubit}

To evaluate the effectiveness of each model in detecting charge jumps across the qubits in
the NEXUS experiment, the efficiency per qubit was calculated. This metric is calculated
as the number of jump-containing signals correctly classified as such over the total number
of signals that contain a jump. The overall efficiency is reported as the mean of the five
per-qubit efficiencies, with the standard deviation across qubits quoted as the error to
characterize qubit-to-qubit variation in model performance.

A comparison of the efficiency results for the baseline models is presented in
Table~\ref{tab:efficiency-float}, while the results following quantization are summarized
in Table~\ref{tab:efficiency-qat}. Similar to the results seen in Sec.~\ref{sec:class-perf},
the tables show that the DCCNN maintains a higher overall efficiency both before and after
quantization by about 4\%. These findings indicate that the ablation of residual
connections was successful, as they do not provide a statistically significant improvement
in detection performance for this task. Consequently, the DCCNN was selected for FPGA
implementation due to its architectural simplicity and reduced computational requirements.

\begin{table}[ht]
    \caption{Per-qubit jump detection efficiency on
        $|\Delta q| \in [0.1, 0.5]\,e$ for the TCN and DCCNN models,
        before quantization.}
    \label{tab:efficiency-float}
    \begin{ruledtabular}
    \begin{tabular}{lcc}
    \textrm{Qubit} & \textrm{TCN} & \textrm{DCCNN} \\
    \colrule
    Q1       & $0.8367$ & $0.8890$ \\
    Q2       & $0.8488$ & $0.9009$ \\
    Q3       & $0.7749$ & $0.8771$ \\
    Q4 (SC)  & $0.6752$ & $0.6881$ \\
    Q4 (SO)  & $0.7626$ & $0.8106$ \\
    \colrule
    Overall  & $0.7797 \pm 0.0621$ & $0.8331 \pm 0.0790$ \\
    \end{tabular}
    \end{ruledtabular}
\end{table}

\begin{table}[ht]
    \caption{Per-qubit jump detection efficiency on
        $|\Delta q| \in [0.1, 0.5]\,e$ for the TCN and DCCNN models after
        quantization-aware training (\texttt{ap\_fixed<16,6>}).}
    \label{tab:efficiency-qat}
    \begin{ruledtabular}
    \begin{tabular}{lcc}
    \textrm{Qubit} & \textrm{TCN} & \textrm{DCCNN} \\
    \colrule
    Q1       & $0.8099$ & $0.8916$ \\
    Q2       & $0.8374$ & $0.9085$ \\
    Q3       & $0.8102$ & $0.8893$ \\
    Q4 (SC)  & $0.6881$ & $0.7060$ \\
    Q4 (SO)  & $0.7601$ & $0.7790$ \\
    \colrule
    Overall  & $0.7811 \pm 0.0528$ & $0.8349 \pm 0.0791$ \\
    \end{tabular}
    \end{ruledtabular}
\end{table}

\subsection{Comparison with \texorpdfstring{$\chi^2$}{chi-squared} algorithm}

To enable a direct comparison with the $\chi^2$ algorithm, we adopt the binned per-qubit
efficiency metric used in Ref.~\cite{Bratrud:2024qnk}. Both the quantized DCCNN and the
$\chi^2$ algorithm are evaluated on simulated datasets of 1,600 scans per qubit with a
jump rate of 1.1~mHz. Jumps are injected at random scan positions with sizes drawn
uniformly from $[0.01, 0.5]\,e$. The simulated scans are produced using the templates and
methodology described in Sec.~\ref{sec:Data-preproc}. Detection efficiency is defined as
the fraction of injected charge jumps that are correctly identified; this differs from the
efficiency calculated in Sec.~\ref{sec:eff-per-qubit}, as the efficiency reported here is
the per-qubit mean efficiency. Statistical uncertainties are estimated as the standard
deviation of the per-qubit mean efficiency across 75 independent sets of simulated scans,
with each set binned in 15 jump-size intervals spanning $|\Delta q| \in [0.1, 0.5]\,e$.

To reproduce the methodology of Ref.~\cite{Bratrud:2024qnk} as closely as possible, the
$\chi^2$ algorithm parameters were tuned on the training dataset described in
Sec.~\ref{sec:training} to maximize jump-tagging efficiency subject to a false-positive-rate
(FPR) constraint of 5\%. This FPR was chosen to match the rates observed for the DCCNN
before and after quantization (6\% and 4\%, respectively).

Table~\ref{tab:efficiency-comparison} presents a side-by-side comparison of the two
methods' per-qubit efficiencies. The DCCNN achieves comparable efficiency to the $\chi^2$
algorithm across all five qubit templates, while requiring no per-qubit parameter tuning
and supporting online deployment at fixed inference latency.

\begin{table}[ht]
    \caption{Comparison of the DCCNN model presented in this work with the $\chi^2$
        algorithm of Ref.~\cite{Bratrud:2024qnk}, showing per-qubit charge jump detection
        efficiency on $|\Delta q| \in [0.1, 0.5]\,e$, evaluated using the methodology of
        Ref.~\cite{Bratrud:2024qnk}.}
    \label{tab:efficiency-comparison}
    \begin{ruledtabular}
    \begin{tabular}{lcc}
    \textrm{Qubit} & $\chi^2$ & \textrm{DCCNN} \\
    \colrule
    Q1       & $0.909 \pm 0.015$ & $0.890 \pm 0.013$ \\
    Q2       & $0.900 \pm 0.015$ & $0.904 \pm 0.015$ \\
    Q3       & $0.885 \pm 0.013$ & $0.890 \pm 0.014$ \\
    Q4 (SC)  & $0.761 \pm 0.018$ & $0.719 \pm 0.021$ \\
    Q4 (SO)  & $0.884 \pm 0.015$ & $0.810 \pm 0.019$ \\
    \colrule
    Mean     & $0.868 \pm 0.007$ & $0.843 \pm 0.022$ \\
    \end{tabular}
    \end{ruledtabular}
\end{table}

\section{Firmware Implementation}

\subsection{The QICK board}

The Quantum Instrumentation Control Kit
(QICK) \cite{qick_github} is an open-source qubit
control platform built on Xilinx RFSoC field-programmable gate arrays
(FPGAs)~\cite{QICK:online}. Originally introduced in 2022~\cite{QICK:2021otz} and
subsequently upgraded to support higher-bandwidth RFSoC chips~\cite{QICK:2023ujm}, QICK
provides a unified hardware and software framework for superconducting qubit control and
readout, requiring only external amplification and filtering.

The QICK firmware architecture centers on a timed processor (tProc) that coordinates
signal generation and readout blocks with precise timing control. Running on the ZCU216
evaluation board, QICK provides 16 DAC channels at 9.85~GS/s and 16 ADC channels at
2.5~GS/s, supporting direct mixer-free signal generation up to 10~GHz and multiplexed
qubit readout. The 100~ps timing resolution enables precisely defined control sequences,
while phase coherence is maintained across all channels without external synchronization.

For charge jump detection, the relevant QICK capability is the integration of custom
signal processing blocks within the FPGA. Recent work has demonstrated the integration of
neural network inference blocks into the QICK readout chain using the \textsc{hls4ml}
workflow~\cite{Duarte:2018ite}, achieving single-shot qubit state discrimination with 96\%
fidelity and 32~ns latency while consuming less than 16\% of the FPGA lookup table
resources~\cite{DiGuglielmo:2025zod}. We adapt this approach for charge jump detection:
rather than classifying single I-Q points for state discrimination, our CNN operates on a
sliding window of Ramsey charge tomography data to identify discontinuous jumps in offset
charge. In that application, the network classifies individual readout samples in the
in-phase/quadrature (I-Q) plane, where the qubit's $|0\rangle$ and $|1\rangle$ states
produce distinguishable points in the complex demodulated signal~\cite{Krantz:2019}.

The QICK platform's combination of high-speed signal generation, low-latency processing,
and programmable logic makes it well-suited for real-time charge jump detection. Upon
detecting a charge jump, the system can trigger immediate qubit recalibration, log the
event timestamp for offline particle physics analysis, or conditionally modify subsequent
control sequences based on the post-jump charge state.

\subsection{Neural network IP integration}

Through the \textsc{hls4ml} workflow, the trained and quantized DCCNN is compiled into a
standalone Intellectual Property (IP) block. In FPGA architecture, an IP block represents
a modular, reusable unit of synthesized logic (VHDL or Verilog) that can be seamlessly
inserted into a larger hardware design \cite{xilinx_ug1037}.

Because the \textsc{hls4ml} streaming backend does not implement convolution dilation,
each dilated layer is lowered to a mathematically identical dilation-1 convolution of
kernel size $(k-1)d+1$ whose kernel is the trained taps interleaved with zeros; since
zero is a fixed point of the quantizer, the network function is unchanged. Quantizer
saturation semantics and accumulator widths are pinned explicitly in the conversion. The
resulting firmware was validated against the QKeras model on 16,000 held-out scans
(99.9\% decision agreement; mean absolute output difference $1.9\times10^{-3}$), and
re-evaluating the full protocol of Table~\ref{tab:efficiency-comparison} through the
compiled firmware reproduces every per-qubit efficiency to within $0.002$.

To integrate this custom machine learning block into the existing QICK ecosystem, we
utilize the Xilinx Advanced eXtensible Interface (AXI) standard \cite{arm_amba_axi}, which
governs data routing across the RFSoC. The integration relies on two primary pathways to
balance high-speed data ingestion with system-level control: AXI4-Stream and AXI4-Lite.

The digitized and decimated time-series data originating from the qubit readout chain is
routed directly into the NN IP via an AXI4-Stream interface. AXI4-Stream is optimized for
high-throughput, latency-critical, and continuous data flow without requiring memory
addresses \cite{xilinx_ug1037}. This makes it the ideal transport mechanism for feeding
the sequential sliding window of the DCCNN. The stream tap is isolated from the readout
chain by an always-ready gate, so back-pressure from the network can never stall data
acquisition; samples arriving while an inference is in flight are dropped. The IP itself
runs in the 100~MHz processing-system clock domain, crossed from the ADC clock by an
AXI4-Stream clock converter, decoupling the network's logic depth from the
timing-critical RF data-converter domains.

Simultaneously, the NN IP connects to the broader system bus and QICK's timed processor
(tProc) \cite{QICK:2021otz} via an AXI4-Lite interface. AXI4-Lite is a memory-mapped
protocol used for low-throughput control and status monitoring. Once the DCCNN computes an
inference, the output (e.g., the jump probability or binary classification flag) is
streamed into an AXI4-Stream FIFO whose contents are accessible to the tProc and
processing system via AXI4-Lite. The tProc can continuously poll this interface; upon
detecting a positive charge jump flag, the tProc can dynamically branch its
instruction sequence in real time, pausing the ongoing experiment to trigger an autonomous
recalibration routine.

\subsection{HLS and resource estimates}

\begin{table*}[ht!]
    \caption{Overall FPGA resources available on Zynq UltraScale+ RFSoC ZCU216 and
        resource utilization for QICK and its ML-enhanced version (post-place-and-route,
        Vivado 2023.1). The memory resources are flip-flops (FF) and block RAMs (BRAM
        36K); the computational resources are look-up tables (LUT) and digital signal
        processors (DSP). QICK baseline from Di~Guglielmo et~al.~\cite{DiGuglielmo:2025zod}.}
    \label{tab:fpga-resources}
    \begin{ruledtabular}
    \begin{tabular}{lcccc}
     & \multicolumn{2}{c}{Memory resources} & \multicolumn{2}{c}{Computational resources} \\
     & FF & BRAM & LUT & DSP \\
    \colrule
    ZCU216       & 850,560 & 1,080 & 425,280 & 4,272 \\
    QICK         & 89,783 (10.56\%) & 309 (28.61\%) & 60,057 (14.12\%) & 481 (11.26\%) \\
    QICK+DCCNN   & 359,030 (42.21\%) & 949 (87.87\%) & 207,244 (48.73\%) & 2,350 (55.01\%) \\
    DCCNN        & $+31.66\%$ & $+59.26\%$ & $+34.61\%$ & $+43.75\%$ \\
    \end{tabular}
    \end{ruledtabular}
\end{table*}

\begin{table}[ht!]
    \caption{Per-layer latency breakdown for the DCCNN IP from Vitis HLS C-synthesis
        (Vivado 2023.1, target clock 10~ns, 100~MHz). Convolution layers report a range
        because the dataflow scheduler resolves their latency depending on input timing;
        non-convolution layers report a single deterministic value.}
    \label{tab:hls-latency}
    \begin{ruledtabular}
    \begin{tabular}{lcc}
    \textrm{Layer} & \textrm{Cycles} & \textrm{Time} \\
    \colrule
    ZeroPad 0           & 138               & $1.38\,\mu$s \\
    Conv1D 0 (dil=1)    & 1,171--2,471      & 11.71--24.71~$\mu$s \\
    ReLU + BN 0         & 261               & $2.61\,\mu$s \\
    ZeroPad 1           & 138               & $1.38\,\mu$s \\
    Conv1D 1 (dil=1)    & 1,171--2,861      & 11.71--28.61~$\mu$s \\
    ReLU + BN 1         & 261               & $2.61\,\mu$s \\
    ZeroPad 2           & 140               & $1.40\,\mu$s \\
    Conv1D 2 (dil=2)    & 1,453--3,169      & 14.53--31.69~$\mu$s \\
    ReLU + BN 2         & 261               & $2.61\,\mu$s \\
    ZeroPad 3           & 140               & $1.40\,\mu$s \\
    Conv1D 3 (dil=2)    & 1,453--3,169      & 14.53--31.69~$\mu$s \\
    ReLU + BN 3         & 261               & $2.61\,\mu$s \\
    AvgPool             & 132               & $1.32\,\mu$s \\
    Dense               & 9--10             & 90--100~ns \\
    Sigmoid             & 2                 & 20~ns \\
    \colrule
    Total latency        & 3,185--3,198      & 31.85--31.98~$\mu$s \\
    Initiation interval  & 1,454--3,170      & 14.54--31.70~$\mu$s \\
    \end{tabular}
    \end{ruledtabular}
\end{table}

Table~\ref{tab:fpga-resources} reports the FPGA resource utilization of the quantized
DCCNN integrated with the QICK platform on the Zynq UltraScale+ RFSoC ZCU216 (Vivado
2023.1). Memory resources are reported as flip-flops (FF), basic memory elements used to
store binary data, and block RAMs (BRAM), dedicated memory blocks that can each store up
to 36~kilobits on the programmable logic. Computational resources are reported as
look-up tables (LUT), configurable logic blocks used for implementing logic functions, and
digital signal processors (DSP), specialized hardware units for efficient computation of
operations such as multiplications and additions. Relative to the QICK baseline, the
DCCNN IP adds $+31.66\%$ FF, $+59.26\%$ BRAM, $+34.61\%$ LUT, and $+43.75\%$ DSP, with
BRAM the tightest constraint at 87.9\% total utilization.

Vitis HLS C-synthesis at a 10~ns target clock (the 100~MHz processing-system clock domain
in which the IP is deployed) estimates a per-inference latency of 3,198 cycles
($32.0\,\mu$s) and an initiation interval of up to 3,170 cycles ($31.7\,\mu$s),
corresponding to a sustained throughput of $\sim 3.2 \times 10^4$ inferences per
second, over four orders of magnitude faster than the millisecond-scale Ramsey scan
cadence. The four causal convolutions dominate the dataflow ($\sim$1,171--3,169 cycles
each; Table~\ref{tab:hls-latency}), while the dense and sigmoid output layers contribute
only 11 cycles. The post-place-and-route design closes timing with positive slack on all
$1.2 \times 10^6$ constrained paths. These results demonstrate that the DCCNN IP can be
integrated into the QICK platform without exhausting resources or introducing prohibitive
latency.

\section{Conclusions}

In this paper, a real-time, FPGA-deployable charge-jump detection system for
superconducting qubits has been presented. This system is based on a 4-layer dilated
causal CNN trained with quantization-aware methods to fixed-point precision
\texttt{ap\_fixed<16,6>}. At only 1,649 parameters, the DCCNN integrates into the QICK
qubit control firmware on the Zynq UltraScale+ ZCU216 with modest resource overhead and
fits comfortably within the available LUT, FF, and DSP budget; BRAM usage of 88\% remains
the binding constraint on further architectural growth.

Evaluated under the methodology of Ref.~\cite{Bratrud:2024qnk}, the DCCNN achieves a mean
per-qubit detection efficiency of $0.843 \pm 0.022$ on $|\Delta q| \in [0.1, 0.5]\,e$,
comparable to the $\chi^2$-window baseline ($0.868 \pm 0.007$) tuned to a matched
false-positive rate. Unlike the $\chi^2$ method, which requires per-qubit hyperparameter
tuning and operates as a stateful per-scan procedure unsuited for real-time inference, the
DCCNN runs as a fixed-latency forward pass without any post-training calibration. A single trained model serves all five qubit templates, eliminating the per-qubit tuning that the classical method requires and thereby improving robustness and reducing analysis bias.

These results show that ML-based charge-jump detection is viable at the resource
and latency constraints of on-chip quantum firmware. The approach is supervision-free at
inference time, requires no hyperparameter tuning when applied to a new qubit, and
supports streaming deployment in the qubit control loop. Future work will focus on
deploying the DCCNN on real-time data acquisition with operating qubit hardware, validating
the simulation-trained model on measured scans and demonstrating in-loop integration with
the QICK control firmware.

\begin{acknowledgments}
The authors from Carnegie Mellon University thank the members of the NEXUS collaboration for providing access to their data and explaining their experimental setup and jump-finding methodology. All authors also acknowledge the work by the authors of Ref. \cite{Bratrud:2024qnk}, which made this real-time ML application possible.

This manuscript has been authored by Fermi Research Alliance, LLC under
Contract No. DE-AC02-07CH11359 with the U.S. Department of Energy, Office of Science, Office
of High Energy Physics.
\end{acknowledgments}

\appendix

\bibliography{ref}

@article{Bratrud:2024qnk,
    author = "Bratrud, G. and others",
    title = "{Measurement of correlated charge noise in superconducting qubits at an underground facility}",
    eprint = "2405.04642",
    archivePrefix = "arXiv",
    primaryClass = "quant-ph",
    reportNumber = "FERMILAB-PUB-24-0199-ETD-PPD",
    doi = "10.1038/s41467-025-63724-4",
    journal = "Nature Commun.",
    volume = "16",
    number = "1",
    pages = "9906",
    year = "2025"
}

@article{Cardani:2022blq,
    author = "Cardani, L. and others",
    title = "{Disentangling the sources of ionizing radiation in superconducting qubits}",
    eprint = "2211.13597",
    archivePrefix = "arXiv",
    primaryClass = "quant-ph",
    reportNumber = "FERMILAB-PUB-22-903-SQMS-TD",
    doi = "10.1140/epjc/s10052-023-11199-2",
    journal = "Eur. Phys. J. C",
    volume = "83",
    number = "1",
    pages = "94",
    year = "2023"
}

@article{Wilen:2020lgg,
    author = "Wilen, C. D. and others",
    title = "{Correlated charge noise and relaxation errors in superconducting qubits}",
    eprint = "2012.06029",
    archivePrefix = "arXiv",
    primaryClass = "quant-ph",
    reportNumber = "FERMILAB-PUB-21-597-PPD",
    doi = "10.1038/s41586-021-03557-5",
    journal = "Nature",
    volume = "594",
    number = "7863",
    pages = "369--373",
    year = "2021"
}

@article{GLASHOW1961579,
title = {Partial-symmetries of weak interactions},
journal = {Nuclear Physics},
volume = {22},
pages = {579},
year = {1961},
issn = {0029-5582},
doi = {10.1016/0029-5582(61)90469-2},
author = {Sheldon L. Glashow}
}

@article{Weinberg:1967tq,
    author = "Weinberg, Steven",
    title = "{A Model of Leptons}",
    doi = "10.1103/PhysRevLett.19.1264",
    journal = "Phys. Rev. Lett.",
    volume = "19",
    pages = "1264",
    year = "1967"
}

@article{Glashow:1970gm,
    author = "Glashow, S. L. and Iliopoulos, J. and Maiani, L.",
    title = "{Weak Interactions with Lepton-Hadron Symmetry}",
    doi = "10.1103/PhysRevD.2.1285",
    journal = "Phys. Rev. D",
    volume = "2",
    pages = "1285",
    year = "1970"
}

@article{GELLMANN1964214,
title = "{A schematic model of baryons and mesons}",
journal = {Phys. Lett.},
volume = {8},
pages = {214},
year = {1964},
issn = {0031-9163},
doi = {10.1016/S0031-9163(64)92001-3},
author = {M. Gell-Mann}
}

@incollection{Zweig:1964jf,
    author = "Zweig, G.",
    editor = "Lichtenberg, D. B. and Rosen, Simon Peter",
    title = "{An SU(3) model for strong interaction symmetry and its breaking. Version 2}",
    booktitle = "{DEVELOPMENTS IN THE QUARK THEORY OF HADRONS. VOL. 1. 1964 - 1978}",
    pages = "22",
    year = "1964",
    doi = "10.17181/CERN-TH-412"
}

@article{DiGuglielmo:2025zod,
    author = "Di Guglielmo, Giuseppe and others",
    title = "{End-to-End Workflow for Machine-Learning-Based Qubit Readout With QICK and hls4ml}",
    eprint = "2501.14663",
    archivePrefix = "arXiv",
    primaryClass = "quant-ph",
    reportNumber = "FERMILAB-PUB-24-0925-ETD-PPD",
    doi = "10.1109/TQE.2025.3604712",
    journal = "IEEE Trans. Quantum Eng.",
    volume = "6",
    pages = "1--10",
    year = "2025"
}

@article{Duarte:2018ite,
    author = "Duarte, Javier and others",
    title = "{Fast inference of deep neural networks in FPGAs for particle physics}",
    eprint = "1804.06913",
    archivePrefix = "arXiv",
    primaryClass = "physics.ins-det",
    reportNumber = "FERMILAB-PUB-18-089-E",
    doi = "10.1088/1748-0221/13/07/P07027",
    journal = "JINST",
    volume = "13",
    number = "07",
    pages = "P07027",
    year = "2018"
}

@misc{QICK:online,
    title = {{QICK: Quantum Instrumentation Control Kit}},
    howpublished = {\url{https://docs.qick.dev}},
    version = {0.2.420},
    note = {Accessed: 2026-01-06}
}

@article{QICK:2021otz,
    author = "Stefanazzi, Leandro and others",
    title = "{The QICK (Quantum Instrumentation Control Kit): Readout and control for qubits and detectors}",
    eprint = "2110.00557",
    archivePrefix = "arXiv",
    primaryClass = "quant-ph",
    reportNumber = "FERMILAB-PUB-21-472-SCD",
    doi = "10.1063/5.0076249",
    journal = "Rev. Sci. Instrum.",
    volume = "93",
    number = "4",
    pages = "044709",
    year = "2022"
}

@article{QICK:2023ujm,
    author = "Ding, Chunyang and others",
    collaboration = "QICK",
    title = "{Experimental advances with the QICK (Quantum Instrumentation Control Kit) for superconducting quantum hardware}",
    eprint = "2311.17171",
    archivePrefix = "arXiv",
    primaryClass = "quant-ph",
    reportNumber = "FERMILAB-PUB-23-783-PPD",
    doi = "10.1103/PhysRevResearch.6.013305",
    journal = "Phys. Rev. Res.",
    volume = "6",
    number = "1",
    pages = "013305",
    year = "2024"
}

@misc{lienhard2021deepneuralnetworkdiscrimination,
      title={Deep Neural Network Discrimination of Multiplexed Superconducting Qubit States}, 
      author={Benjamin Lienhard and Antti Vepsäläinen and Luke C. G. Govia and Cole R. Hoffer and Jack Y. Qiu and Diego Ristè and Matthew Ware and David Kim and Roni Winik and Alexander Melville and Bethany Niedzielski and Jonilyn Yoder and Guilhem J. Ribeill and Thomas A. Ohki and Hari K. Krovi and Terry P. Orlando and Simon Gustavsson and William D. Oliver},
      year={2021},
      eprint={2102.12481},
      archivePrefix={arXiv},
      primaryClass={quant-ph},
      url={https://arxiv.org/abs/2102.12481}, 
}

@article{Navarathna_2021,
   title={Neural networks for on-the-fly single-shot state classification},
   volume={119},
   ISSN={1077-3118},
   url={http://dx.doi.org/10.1063/5.0065011},
   DOI={10.1063/5.0065011},
   number={11},
   journal={Applied Physics Letters},
   publisher={AIP Publishing},
   author={Navarathna, Rohit and Jones, Tyler and Moghaddam, Tina and Kulikov, Anatoly and Beriwal, Rohit and Jerger, Markus and Pakkiam, Prasanna and Fedorov, Arkady},
   year={2021},
   month=sep }

@inproceedings{NIPS2012_c399862d,
 author = {Krizhevsky, Alex and Sutskever, Ilya and Hinton, Geoffrey E},
 booktitle = {Advances in Neural Information Processing Systems},
 editor = {F. Pereira and C.J. Burges and L. Bottou and K. Weinberger},
 pages = {1097–-1105},
 publisher = {Curran Associates, Inc.},
 title = {ImageNet Classification with Deep Convolutional Neural Networks},
 url = {https://proceedings.neurips.cc/paper_files/paper/2012/file/c399862d3b9d6b76c8436e924a68c45b-Paper.pdf},
 volume = {25},
 year = {2012}
}

@misc{jwa2022realtimeinference2dconvolutional,
      title={Real-time Inference with 2D Convolutional Neural Networks on Field Programmable Gate Arrays for High-rate Particle Imaging Detectors}, 
      author={Yeon-jae Jwa and Giuseppe Di Guglielmo and Lukas Arnold and Luca Carloni and Georgia Karagiorgi},
      year={2022},
      eprint={2201.05638},
      archivePrefix={arXiv},
      primaryClass={physics.ins-det},
      url={https://arxiv.org/abs/2201.05638}, 
}

@misc{samaha2026automaticchargestatetuning,
      title={Automatic Charge State Tuning of 300 mm FDSOI Quantum Dots Using Neural Network Segmentation of Charge Stability Diagram}, 
      author={Peter Samaha and Amine Torki and Ysaline Renaud and Sam Fiette and Emmanuel Chanrion and Pierre-Andre Mortemousque and Yann Beilliard},
      year={2026},
      eprint={2604.13662},
      archivePrefix={arXiv},
      primaryClass={cond-mat.mes-hall},
      url={https://arxiv.org/abs/2604.13662}, 
}

@article{Yon_2025,
   title={Experimental Online Quantum Dots Charge Autotuning Using Neural Networks},
   volume={25},
   ISSN={1530-6992},
   url={http://dx.doi.org/10.1021/acs.nanolett.4c04889},
   DOI={10.1021/acs.nanolett.4c04889},
   number={10},
   journal={Nano Letters},
   publisher={American Chemical Society (ACS)},
   author={Yon, Victor and Galaup, Bastien and Rohrbacher, Claude and Rivard, Joffrey and Morel, Alexis and Leclerc, Dominic and Godfrin, Clément and Li, Ruoyu and Kubicek, Stefan and Greve, Kristiaan De and Dupont Ferrier, Eva and Beilliard, Yann and Melko, Roger G. and Drouin, Dominique},
   year={2025},
   month=feb, pages={3717–-3725} }

@article{DBLP:journals/corr/abs-1803-01271,
  author       = {Shaojie Bai and
                  J. Zico Kolter and
                  Vladlen Koltun},
  title        = {An Empirical Evaluation of Generic Convolutional and Recurrent Networks
                  for Sequence Modeling},
  journal      = {CoRR},
  volume       = {abs/1803.01271},
  year         = {2018},
  url          = {http://arxiv.org/abs/1803.01271},
  eprinttype   = {arXiv},
  eprint       = {1803.01271},
  bibsource    = {dblp computer science bibliography, https://dblp.org}
}

@misc{chollet2015keras,
  title={Keras},
  author={Chollet, Fran\c{c}ois and others},
  year={2015},
  howpublished={\url{https://keras.io}},
}

@article{coelho2021qkeras,
  author  = {Coelho, Claudionor N. and Kuusela, Aki and Li, Shan and Zhuang, Hao and Ngadiuba, Jennifer and Aarrestad, Thea Klaeboe and Loncar, Vladimir and Pierini, Maurizio and Pol, Adrian Alan and Summers, Sioni},
  title   = {Automatic heterogeneous quantization of deep neural networks for low-latency inference on the edge for particle detectors},
  journal = {Nature Machine Intelligence},
  volume  = {3},
  number  = {8},
  pages   = {675--686},
  year    = {2021},
  doi     = {10.1038/s42256-021-00356-5}
}

@misc{qkeras_software,
  author       = {Coelho, Claudionor N. and others},
  title        = {{QKeras}: a quantization deep learning library for {Tensorflow Keras}},
  howpublished = {\url{https://github.com/google/qkeras}},
  year         = {2019}
}

@misc{jacob2017quantization,
  title         = {Quantization and Training of Neural Networks for Efficient Integer-Arithmetic-Only Inference}, 
  author        = {Benoit Jacob and Skirmantas Kligys and Bo Chen and Menglong Zhu and Matthew Tang and Andrew Howard and Hartwig Adam and Dmitry Kalenichenko},
  year          = {2017},
  eprint        = {1712.05877},
  archivePrefix = {arXiv},
  primaryClass  = {cs.LG}
}

@inproceedings{Chen2020Statistical,
  author    = {Chen, J. and Gai, Y. and Yao, Z. and Mahoney, M. W. and Gonzalez, J. E.},
  title     = {A Statistical Framework for Low-bitwidth Training of Deep Neural Networks},
  booktitle = {Advances in Neural Information Processing Systems (NeurIPS)},
  volume    = {33},
  pages     = {1823--1834},
  year      = {2020},
  publisher = {Curran Associates, Inc.}
}

@inproceedings{Chen2021ActNN,
  author    = {Chen, J. and Zheng, L. and Yao, Z. and Wang, D. and Stoica, I. and Mahoney, M. W. and Gonzalez, J. E.},
  title     = {ActNN: Reducing Training Memory Footprint via 2-Bit Activation Compressed Training},
  booktitle = {Proceedings of the 38th International Conference on Machine Learning (ICML)},
  series    = {Proceedings of Machine Learning Research},
  volume    = {139},
  pages     = {1803--1813},
  year      = {2021},
  publisher = {PMLR}
}

@manual{xilinx_ug1037,
  title        = {{AXI Reference Guide (UG1037)}},
  organization = {Xilinx Inc.},
  year         = {2017},
  note         = {Version 4.0}
}

@manual{arm_amba_axi,
  title        = {{AMBA AXI and ACE Protocol Specification}},
  organization = {ARM Ltd.},
  year         = {2011},
  note         = {ARM IHI 0022E}
}

@misc{danilin2024quantumsensingtunablesuperconducting,
      title={Quantum sensing with tunable superconducting qubits: optimization and speed-up}, 
      author={Sergey Danilin and Nicholas Nugent and Martin Weides},
      year={2024},
      eprint={2211.08344},
      archivePrefix={arXiv},
      primaryClass={quant-ph},
      url={https://arxiv.org/abs/2211.08344}, 
}

@article{Bonomo:2026num,
    author = "Bonomo, Camilla and others",
    title = "{First operation of superconducting quantum bits as particle detectors}",
    reportNumber = "FERMILAB-CONF-26-0260-SQMS-TD",
    doi = "10.22323/1.519.0001",
    journal = "PoS",
    volume = "14YRM2025",
    pages = "001",
    year = "2026"
}

@article{Ramanathan_2026,
   title={Quantum parity detectors: A qubit-based particle-detection scheme with meV thresholds for rare-event searches},
   volume={1},
   ISSN={3070-2240},
   url={http://dx.doi.org/10.1103/kqd2-spb1},
   DOI={10.1103/kqd2-spb1},
   journal={APS Open Science},
   publisher={American Physical Society (APS)},
   author={Ramanathan, K. and Sandoval, B. J. and Parker, J. E. and Joshi, L. M. and Beyer, A. D. and Echternach, P. M. and Rosenblum, S. and Golwala, S. R.},
   year={2026},
   month=May }

@misc{qick_github,
    title = {{QICK}: {Q}uantum {I}nstrumentation {C}ontrol {K}it},
    author = {{Fermi National Accelerator Laboratory}},
    howpublished = {\url{https://github.com/openquantumhardware/qick}},
    version = {0.2.420},
    note = {Open-source software, accessed 2026-07-15},
    year = {2024}
}

@article{Krantz:2019,
  author  = {Krantz, P. and Kjaergaard, M. and Yan, F. and Orlando, T. P.
              and Gustavsson, S. and Oliver, W. D.},
  title   = {A Quantum Engineer's Guide to Superconducting Qubits},
  journal = {Applied Physics Reviews},
  volume  = {6},
  number  = {2},
  pages   = {021318},
  year    = {2019},
  doi     = {10.1063/1.5089550}
}

@article{Dixit_2021,
   title={Searching for Dark Matter with a Superconducting Qubit},
   volume={126},
   ISSN={1079-7114},
   url={http://dx.doi.org/10.1103/PhysRevLett.126.141302},
   DOI={10.1103/physrevlett.126.141302},
   number={14},
   journal={Physical Review Letters},
   publisher={American Physical Society (APS)},
   author={Dixit, Akash V. and Chakram, Srivatsan and He, Kevin and Agrawal, Ankur and Naik, Ravi K. and Schuster, David I. and Chou, Aaron},
   year={2021},
   month=Apr }

@article{Fink_2024,
   title={Superconducting quasiparticle-amplifying transmon: A qubit-based sensor for meV-scale phonons and single terahertz photons},
   volume={22},
   ISSN={2331-7019},
   url={http://dx.doi.org/10.1103/PhysRevApplied.22.054009},
   DOI={10.1103/physrevapplied.22.054009},
   number={5},
   journal={Physical Review Applied},
   publisher={American Physical Society (APS)},
   author={Fink, C.W. and Salemi, C.P. and Young, B.A. and Schuster, D.I. and Kurinsky, N.A.},
   year={2024},
   month=Nov }

@article{Linehan_2025,
   title={Estimating the energy threshold of phonon-mediated superconducting qubit detectors operated in an energy-relaxation sensing scheme},
   volume={111},
   ISSN={2470-0029},
   url={http://dx.doi.org/10.1103/PhysRevD.111.063047},
   DOI={10.1103/physrevd.111.063047},
   number={6},
   journal={Physical Review D},
   publisher={American Physical Society (APS)},
   author={Linehan, R. and Hernandez, I. and Temples, D. J. and Dang, S. Q. and Baxter, D. and Hsu, L. and Figueroa-Feliciano, E. and Khatiwada, R. and Anyang, K. and Bowring, D. and Bratrud, G. and Cancelo, G. and Chou, A. and Gualtieri, R. and Stifter, K. and Sussman, S.},
   year={2025},
   month=Mar }

@misc{linehan2024listeningnewphysicsquantum,
      title={Listening For New Physics With Quantum Acoustics}, 
      author={Ryan Linehan and Tanner Trickle and Christopher R. Conner and Sohitri Ghosh and Tongyan Lin and Mukul Sholapurkar and Andrew N. Cleland},
      year={2024},
      eprint={2410.17308},
      archivePrefix={arXiv},
      primaryClass={hep-ph},
      url={https://arxiv.org/abs/2410.17308}, 
}

@misc{celi2026measuringquasiparticledynamicsparticle,
      title={Measuring quasiparticle dynamics for particle impact reconstruction in a superconducting qubit chip}, 
      author={E. Celi and R. Linehan and P. M. Harrington and M. Li and H. D. Pinckney and K. Serniak and W. D. Oliver and J. A. Formaggio and E. Figueroa-Feliciano and D. Baxter},
      year={2026},
      eprint={2604.13176},
      archivePrefix={arXiv},
      primaryClass={quant-ph},
      url={https://arxiv.org/abs/2604.13176}, 
}

@article{Thorbeck_2023,
   title={Two-Level-System Dynamics in a Superconducting Qubit Due to Background Ionizing Radiation},
   volume={4},
   ISSN={2691-3399},
   url={http://dx.doi.org/10.1103/PRXQuantum.4.020356},
   DOI={10.1103/prxquantum.4.020356},
   number={2},
   journal={PRX Quantum},
   publisher={American Physical Society (APS)},
   author={Thorbeck, Ted and Eddins, Andrew and Lauer, Isaac and McClure, Douglas T. and Carroll, Malcolm},
   year={2023},
   month=jun }

\end{document}